\documentclass[sigconf]{acmart}
\usepackage{enumitem}
\usepackage{subfigure}

\AtBeginDocument{%
  \providecommand\BibTeX{{%
    \normalfont B\kern-0.5em{\scshape i\kern-0.25em b}\kern-0.8em\TeX}}}

\setcopyright{acmlicensed}
\copyrightyear{2024}
\acmYear{2024}
\setcopyright{rightsretained}
\acmConference[ICCAD '24]{IEEE/ACM International Conference on
Computer-Aided Design}{October 27--31, 2024}{New York, NY, USA}
\acmBooktitle{IEEE/ACM International Conference on Computer-Aided Design
(ICCAD '24), October 27--31, 2024, New York, NY, USA}
\acmDOI{10.1145/3676536.3676811}
\acmISBN{979-8-4007-1077-3/24/10}

\graphicspath{{./images/}}

\begin{document}

\title{Detecting Fraudulent Services on Quantum Cloud Platforms via Dynamic Fingerprinting}

\author{Jindi Wu}
\email{jwu21@wm.edu}
\affiliation{%
  \institution{William \& Mary}
 \city{Williamsburg}
 \state{Virginia}
 \country{USA}
}

\author{Tianjie Hu}
\email{thu04@wm.edu}
\affiliation{%
  \institution{William \& Mary}
 \city{Williamsburg}
 \state{Virginia}
 \country{USA}
}

\author{Qun Li}
\email{liqun@cs.wm.edu}
\affiliation{%
  \institution{William \& Mary}
 \city{Williamsburg}
 \state{Virginia}
 \country{USA}
}

\renewcommand{\shortauthors}{Wu et al.}

\begin{abstract}
Noisy Intermediate-Scale Quantum (NISQ) devices, while accessible via cloud platforms, face challenges due to limited availability and suboptimal quality. These challenges raise the risk of cloud providers offering fraudulent services. This emphasizes the need for users to detect such fraud to protect their investments and ensure computational integrity. This study introduces a novel dynamic fingerprinting method for detecting fraudulent service provision on quantum cloud platforms, specifically targeting machine substitution and profile fabrication attacks. The dynamic fingerprint is constructed using a \textit{single} probing circuit to capture the unique error characteristics of quantum devices, making this approach practical because of its trivial computational costs. When the user examines the service, the execution results of the probing circuit act as the device-side fingerprint of the quantum device providing the service. The user then generates the user-side fingerprint by estimating the expected execution result, assuming the correct device is in use. We propose an algorithm for users to construct the user-side fingerprint with linear complexity. By comparing the device-side and user-side fingerprints, users can effectively detect fraudulent services. Our experiments on the IBM Quantum platform, involving seven devices with varying capabilities, confirm the method's effectiveness. 
% We demonstrate that a Manhattan distance exceeding 0.035 between two fingerprints indicates a fraudulent service has been detected.
\end{abstract}

%%
%% The code below is generated by the tool at: http://dl.acm.org/ccs.cfm






%% The following are not a requirement, delete if not using
% \received{20 February 2024}  %% inital submission date
% \received[revised]{12 March 2024} %% interim new draft
% \received[accepted]{5 June 2024}  %% publication version

\maketitle

\section{Introduction}

Currently, the NISQ resources are readily accessible through cloud-based platforms, enabling users to harness quantum devices via the internet without the need for physical ownership \cite{preskill2018quantum, singh2014quantum, das2019case}. Many companies now offer quantum cloud platforms, broadening the reach of quantum computing. Notable platforms like Microsoft Azure Quantum \cite{Azure} and Amazon Braket \cite{AmazonBraket} offer access to real quantum hardware from multiple providers such as Quantinuum, IonQ, Rigetti, Pasqal, Quantum Circuits, etc., through their cloud services. Users are typically charged based on either machine time or the types of gates used in their circuits. For example, IBM Quantum Platform's Pay-As-You-Go plan provides access to advanced processors for \$1.60 per quantum second \cite{IBMQ, castelvecchi2017ibm}, while IonQ machines on Azure charge based on quantum operations, starting at \$97.50 for a program that includes error mitigation.

\begin{figure}[t!]
\centering 
\includegraphics[width=1\linewidth]{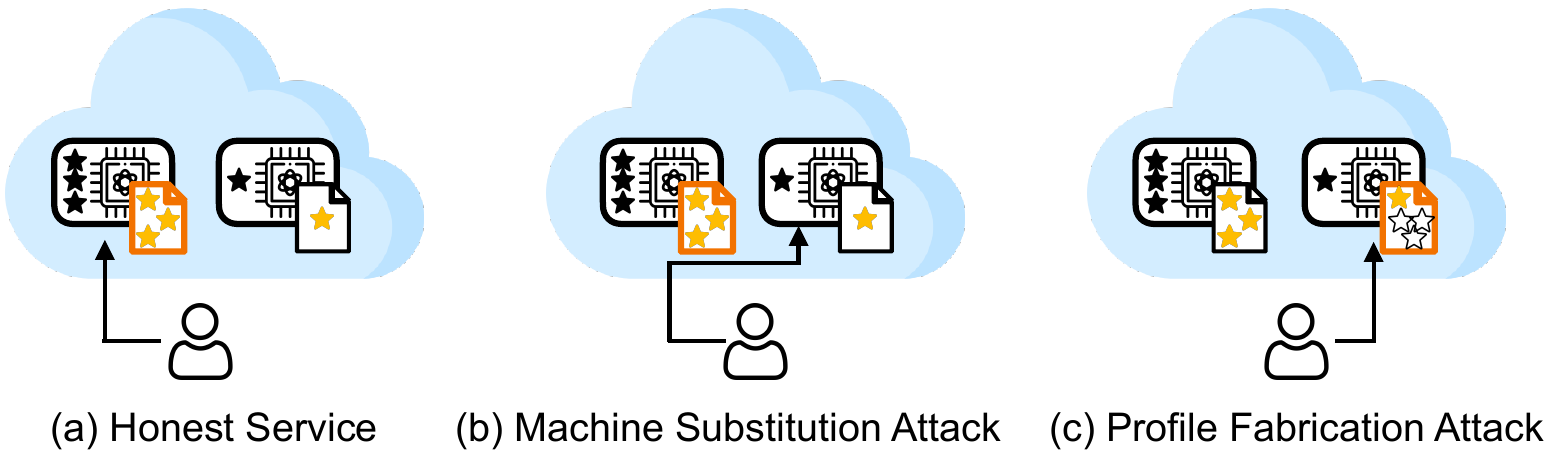}
\caption{\textbf{Honest and fraudulent services.} (a) Honest Service: Users select a device based on its profile, and the user's job is executed on the selected device as intended. (b) Machine Substitution Attack: Users select a high-quality device based on its profile, but the cloud platform executes the user's job on a substituted device of lower quality. (c) Profile Fabrication Attack: The cloud provider fabricates the profile of a device to falsely indicate improved quality, tricking users into selecting this misrepresented device.}
\label{fig:threat}
\end{figure}

Manufacturing imperfections in quantum computers lead to varying quality across machines. Consequently, NISQ machines on these cloud platforms come with a profile that characterizes each machine, describing the qubit topology, the stability of qubits, and the error rates of quantum operations \cite{ravi2021quantum}. 
Typically, users select quantum devices carefully based on these profiles, aiming to minimize costs and meet specific requirements. For instance, a user seeking high fidelity in circuit execution might choose a quantum device with the lowest error rates for the required quantum operators.

However, as quantum computing cloud platforms rapidly evolve, concerns about their trustworthiness have emerged \cite{phalak2021quantum, ash2020analysis, upadhyay2023obfuscating, upadhyay2023trustworthy, xu2023exploration, mi2022securing}. 
On these platforms, users often prefer high-quality, powerful quantum devices, leaving lower-quality machines idle. This raises significant concerns about resource allocation integrity, as cloud providers might offer fraudulent services to maximize platform throughput and profitability.
As shown in Fig.~\ref{fig:threat}, this integrity can be compromised if a cloud provider secretly allocates a quantum device not chosen by the user, potentially failing to meet quality requirements or leaking private information \cite{rahaman2015review}. Furthermore, fabricating a device's profile to present higher quality than actual performance may mislead users into selecting inferior machines.
Such compromised integrity may lead to a significant discrepancy between the expected and actual performance of users' quantum circuits. 
Therefore, it is crucial for users to verify the transparency of hardware allocation and the accuracy of the device's profile to protect their interests.

Fingerprinting techniques are used to uniquely identify quantum devices or detect inconsistencies between device profiles and their actual performance. Essentially, a device-side fingerprint is created using the unique characteristics of the quantum device. Users then access the device to collect these characteristics and construct a user-side fingerprint. Finally, by comparing the device-side and user-side fingerprints, fraudulent services can be identified.
Existing error-based quantum device fingerprinting methods have successfully utilized quantum errors as unique identifiers for quantum devices \cite{mi2021short, martina2022learning}. However, they are impractical for current quantum cloud platforms due to the high costs involved in generating and updating fingerprints. These challenges stem from the inherent complexity and instability of quantum errors.

In this paper, we introduce a practical quantum error-based dynamic fingerprinting approach that can verify whether a fraudulent service is provided. The proposed dynamic fingerprinting approach capitalizes on the diversity of quantum errors without being hindered by their instability. We demonstrate that the noisy execution result of a \textit{single} probing circuit effectively captures the unique characteristics of a quantum device and can serve as the fingerprint for quantum devices. With this designed device-side fingerprint, the user-side fingerprint is an estimated execution result for the probing circuit, assuming it is executed on the target device.
To facilitate this, we introduce a \textit{qubit survivability estimation algorithm} that predicts the execution result of the probing circuit. If there is a mismatch between the device-side and user-side fingerprints, it likely indicates the provision of a fraudulent service. Our approach is highly efficient in utilizing quantum resources, as it requires the execution of a single probing circuit. The overhead involved in estimating the results for this probing circuit is trivial, scaling linearly with the circuit's size. 
We evaluate the proposed dynamic fingerprinting approach using seven quantum devices of varying sizes on the IBM Quantum platform. The results indicate that fraudulent services can be detected by setting a threshold of 0.035 for the Manhattan distance between the device-side and user-side fingerprints.

The primary contributions of our work are encapsulated in three main points:
\begin{itemize}[leftmargin=*]
    \item To the best of the authors' knowledge, this is the first approach to introducing a dynamic fingerprint for quantum devices that capitalizes on the inherent diversity and instability of quantum errors.
    \item We analyze the fine-grained error accumulation within specific quantum circuits, offering a lightweight and reliable method for estimating their noisy execution results.
    \item We conducted extensive experiments on an actual quantum cloud platform, demonstrating the effectiveness and advantages of our proposed approach.
\end{itemize}

\section{Related Work} \label{sec:related}

Several approaches have been proposed to use quantum errors as distinctive features for fingerprinting \cite{martina2022learning, mi2021short}. In \cite{martina2022learning}, carefully designed testbed circuits are executed multiple times on different quantum devices, and the noisy results are used to train a machine learning (ML) classifier to capture distinct error patterns of devices. This classifier can then verify whether a specific quantum device executed the probing circuit. However, $n$ binary classifiers are required for $n$ involved quantum devices, making the cost considerably high.
Alternatively, the work \cite{mi2021short} aims to create unique fingerprints based on crosstalk errors by training an ML model using the noisy results from a sequence of probing circuits to capture crosstalk error patterns on a quantum device.
Although these methods can identify cloud-based quantum machines and detect substitution attacks, they demand significant quantum and classical resources and require retraining after each machine calibration, making them impractical for current quantum cloud platforms. Furthermore, they cannot detect profile fabrication attacks. To address these limitations, we propose an efficient error-based approach for detecting counterfeit quantum hardware, capable of identifying both machine substitution and profile fabrication attacks. This method is highly practical for implementation on current quantum cloud platforms, requiring trivial quantum and classical resources

Other characteristics of quantum devices have been explored for identifying quantum machines. The work \cite{morris2023fingerprinting} investigates the use of SRAM-based physically unclonable functions (PUFs) to generate digital fingerprints for quantum devices under cryogenic conditions, offering a unique and secure method. Moreover, \cite{smith2023fast} proposes a fingerprinting method based on qubit frequencies for fixed-frequency transmon qubits, claiming these frequencies can serve as reliable identifiers across different quantum machines over time. However, these methods are not accessible to users relying on current quantum cloud platforms for detecting fraudulent services.

\section{Threat Model} \label{sec:threat}

\begin{figure*}[t]
\centering 
\includegraphics[width=0.95\textwidth]{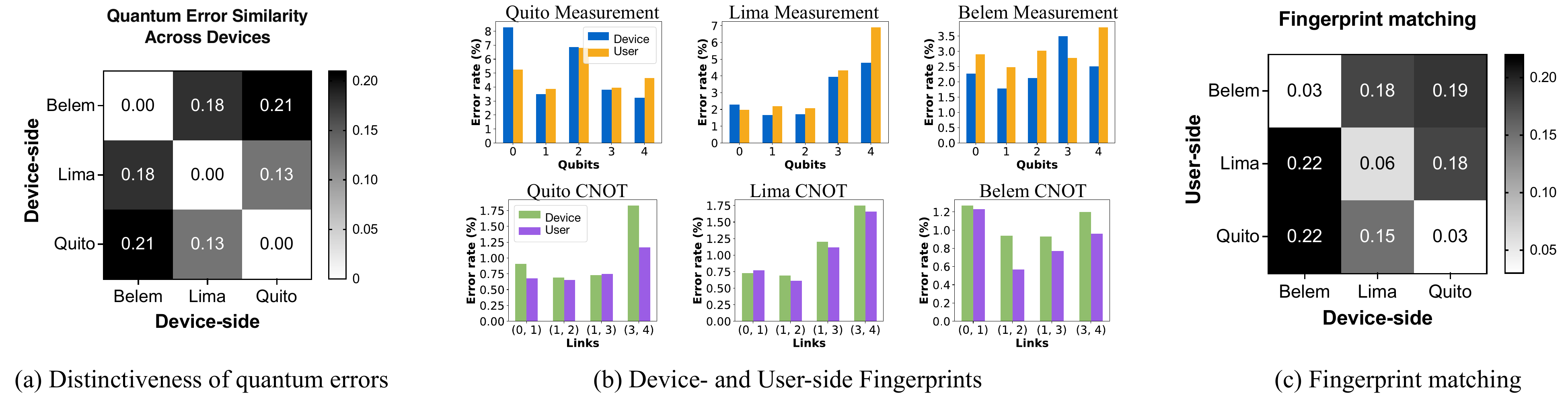}
\caption{\textbf{Basic Quantum Error-Based Fingerprinting Approach.} }
\label{fig:cd_fingerprint}
\end{figure*}

Quantum cloud platforms facilitate services by managing job queues for available devices. Initially, users design circuits tailored to their specific problems. These circuits are then transpiled into executable formats compatible with the characteristics of the selected quantum device, as outlined in its device profile. This profile, detailing qubit information, qubit topology, and operational errors, is updated daily following device calibration. The transpilation process includes converting complex quantum gates to basic types supported by the device, routing qubits in alignment with the qubit topology, and optimizing gate placement according to error rates. After transpilation, users package these circuits into jobs and submit them to the job queue of the chosen quantum device, where they await execution. 
However, if the transpiled circuit does not align with the qubit topology of the quantum device, the cloud platform will report an error to the user.

In our threat model, we assume that a malicious cloud provider aims to enhance platform throughput through two primary attacks: machine substitution and profile fabrication. We assume that the cloud provider does not modify the user-submitted circuits and only has the capability to execute them. Therefore, to successfully run a user's circuit without reporting errors, the provider may substitute the selected device with another quantum device that has the same qubit topology for the involved qubits.
We assume that the platform executes the user's quantum circuits on actual quantum devices, rather than simulating them. We assume that the user has access to the profiles of all available quantum devices but cannot modify them.
In a machine substitution attack, the provider might assign a lower-quality or underutilized quantum device to a user's job, as shown in Fig.~\ref{fig:threat} (b). This unauthorized execution, typically without user consent, can lead to compromised results and negatively impact user interests. 
Moreover, in a profile fabrication attack, the malicious cloud provider misled users into using relatively idle quantum machines by fabricating machine profiles to display falsely low error rates for quantum operations, thereby presenting these machines as higher quality than they actually are, as shown in Fig.~\ref{fig:threat} (c).

\section{Quantum Errors as Fingerprint} \label{sec:basic}

Before presenting the proposed approach, we first introduce a straightforward method of using quantum errors as a device fingerprint, discussing both its advantages and disadvantages.

\textbf{Distinctiveness.} First, we highlight the distinctiveness of quantum errors. While previous methods have successfully utilized ML techniques to capture the unique error patterns of quantum devices, we explore the efficacy of quantum errors themselves in distinguishing between devices. We examine three similar quantum devices on the IBM Quantum cloud, each sharing the same 5-qubit topology and gate set. For each device, we construct an error vector containing error rates associated with operations, such as CNOT gates and measurements. 
The error vector is represented as \(E = \{CNOT_{(0, 1)}: e_1, CNOT_{(1, 2)}: e_2, \ldots, Meas_0: e_j, Meas_1: e_{j+1}, \ldots\}\), with a total length of 9 for the devices considered. The error rates are documented in the device profiles, and accessible to users.
To compare device error vectors, we compute the Manhattan distance between each pair of vectors. A smaller distance indicates stronger matching. Fig.~\ref{fig:cd_fingerprint} (a) illustrates that a device's error vector closely matches itself and notably differs from others', confirming that quantum error is a distinctive characteristic of quantum devices.

\begin{figure*}[t]
\centering 
\includegraphics[width=1\textwidth]{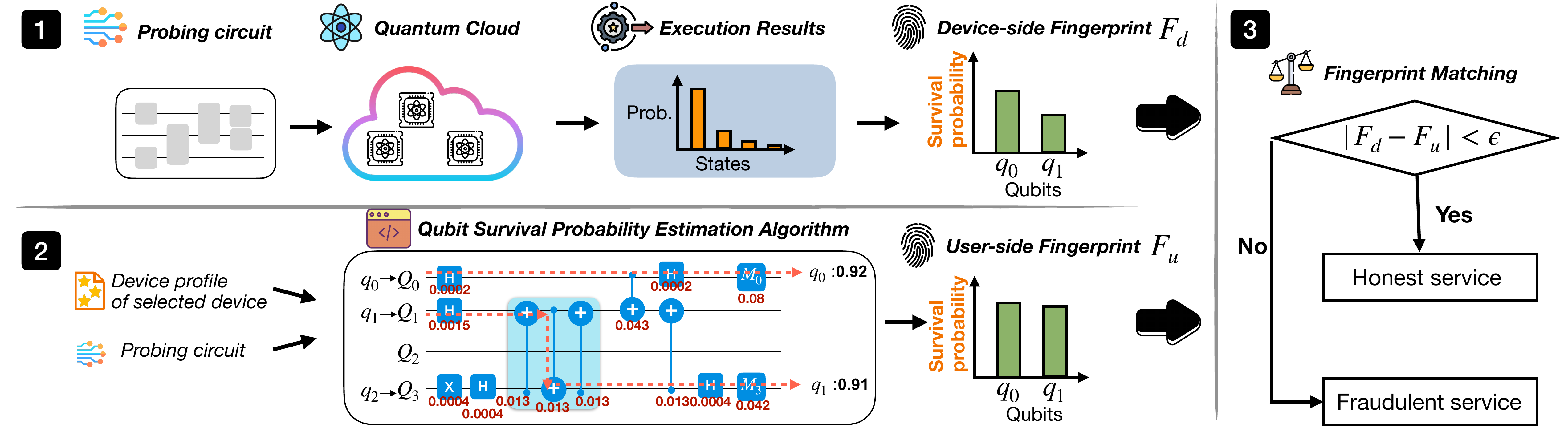}
\caption{Overview of Dynamic Error-Based Fingerprinting. Step 1: Device-side fingerprint generation; Step 2: User-side fingerprint generation; Step 3: Fingerprint matching.}
\label{fig:overview}
\end{figure*}

\textbf{Effectiveness.} Next, we implement a basic error-based fingerprinting to identify quantum devices, assuming that the profiles of these devices are authentic.
We define the device-side fingerprint as the error vector $E$ constructed by retrieving the error rates of operations from the device profiles. These error rates are determined during hardware calibration, which includes calculating CNOT error rates through randomized benchmarking \cite{knill2008randomized, magesan3639robust, magesan2012characterizing} and assessing measurement errors using state preparation and measurement techniques. Similarly, the user-side fingerprint is generated using the same methods. Fig.~\ref{fig:cd_fingerprint} (b) details the device-side and user-side fingerprints. To identify the quantum devices, we perform fingerprint matching by calculating the Manhattan distance between fingerprints, as illustrated in Fig.~\ref{fig:cd_fingerprint} (c). It demonstrates that the user-side fingerprint of a quantum device only strongly matches its corresponding device-side fingerprint and distinctly differs from others. This basic approach shows the effectiveness of using quantum errors themselves as unique identifiers of quantum devices.

\textbf{Temporary stability.} Additionally, the effectiveness of this approach demonstrates the temporary stability of quantum errors. Specifically, in Fig.~\ref{fig:cd_fingerprint} (b), the error rates were calculated by the user 7, 23, and 1 hour(s) after device calibration, respectively. Although some error rates show deviations from those calculated during calibration, most error rates remain similar in value and do not compromise the differentiation of devices. Consequently, we conclude that quantum error rates exhibit temporary stability within the same calibration cycle.

\textbf{Scalability.} Nevertheless, this approach is \textit{not scalable} because it requires substantial quantum computing resources. Specifically, calculating a CNOT error rate involves running 60 randomized benchmarking circuits, and constructing a fingerprint for each quantum device requires a total of 270 circuits. The number of circuits needed increases significantly for larger quantum devices. While one might attempt to reduce costs by selectively calculating and comparing error rates of a subset of CNOT gates and measurement errors, the expense of conducting multiple batches of randomized benchmarking circuits remains unavoidable. Additionally, since device error profiles are updated regularly, frequently updating fingerprints is crucial to maintain accurate and reliable device identification.

In summary, while the quantum error-based fingerprinting approach is effective, its high cost and scalability issues limit its practicality on current quantum cloud platforms. To address this, we propose dynamic fingerprinting as an efficient alternative solution.

\section{Dynamic Fingerprinting Approach} \label{sec:advanced}

We present a lightweight dynamic fingerprinting method for detecting fraudulent services on quantum cloud platforms. To improve computational efficiency, we design \textit{dynamic} device fingerprints based on a dynamic characteristic—quantum errors. 
Unlike previous error-based fingerprinting methods that create static fingerprints from dynamic errors and require multiple circuits to collect device characteristics, as well as costly, frequent updates after each calibration, our approach incorporates the impact of quantum errors into the probing circuit's execution results. By integrating fingerprint generation with user access, we reduce the cost of fingerprint creation and eliminate the need for fingerprint updates. 
Our method requires only a \textit{single} probing circuit for each detection, significantly reducing computational costs. The overview of this approach, shown in Fig.~\ref{fig:overview}, consists of three main steps:
\begin{itemize}[leftmargin=*]
    \item \textbf{Device-side Fingerprint Generation:} Users submit a designed probing circuit to the quantum cloud platform and use its execution results as the fingerprint of the quantum device that provides service.
    \item \textbf{User-side Fingerprint Generation:} Users estimate the probing circuit's execution result as the user-side fingerprint, assuming it is run on the selected quantum device.
    \item \textbf{Fingerprint Matching:} Users compare the device-side and user-side fingerprints to identify potential fraudulent services. The average qubit Manhattan distance is used to measure the difference between fingerprints. A distance exceeding 0.035 is considered a sign of a fraudulent service.
\end{itemize}

\subsection{Device-side Fingerprint}

\subsubsection{Distinctiveness of circuit outputs}
\begin{figure}[t]
\centering 
\includegraphics[width=0.95\linewidth]{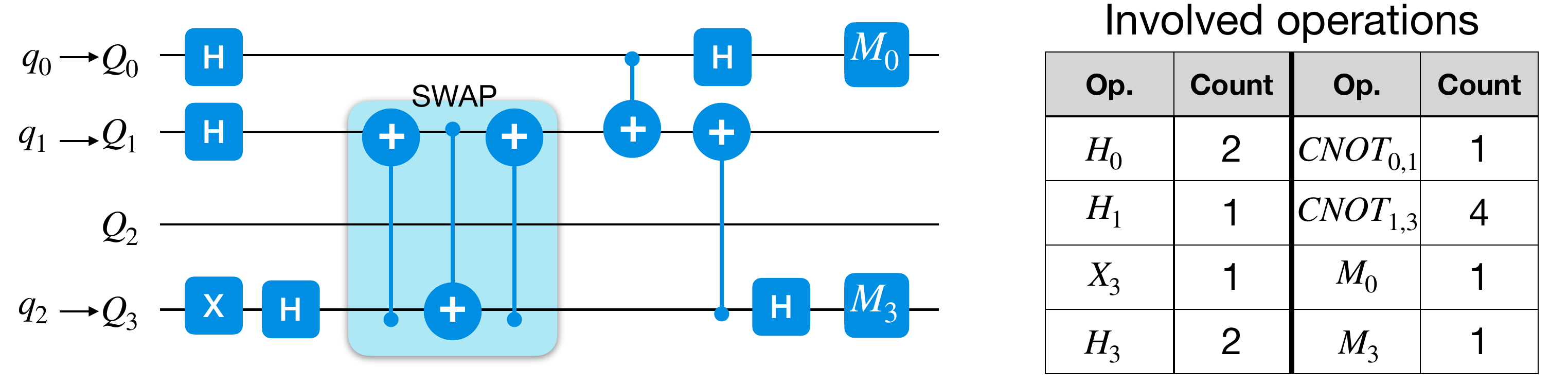}
\caption{Example of a Probing Circuit. Left: a three-qubit BV circuit transpiled with a qubit mapping of [0, 1, 3]. Right: the quantum operations involved in the circuit.}
\label{fig:prob_circ}
\end{figure}

\begin{figure}[t!]
    \centering
    \subfigure[]{
    \label{fig:circ}
    \includegraphics[width=.41\linewidth]{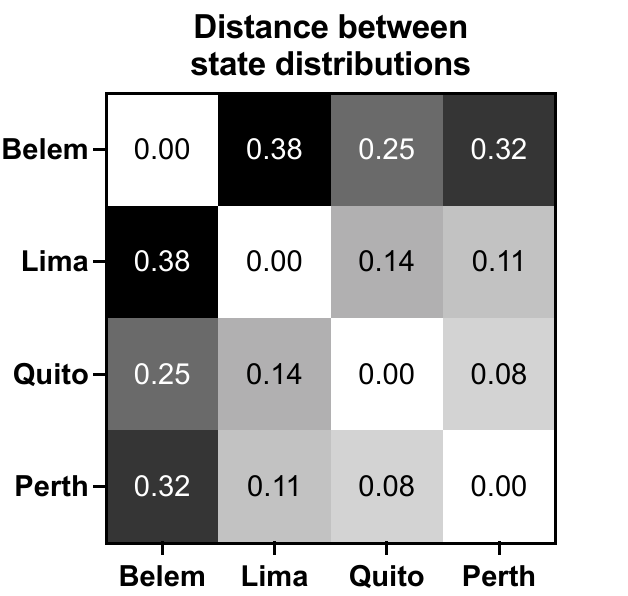}
    }
    \subfigure[]{
    \label{fig:qubit}
    \includegraphics[width=.41\linewidth]{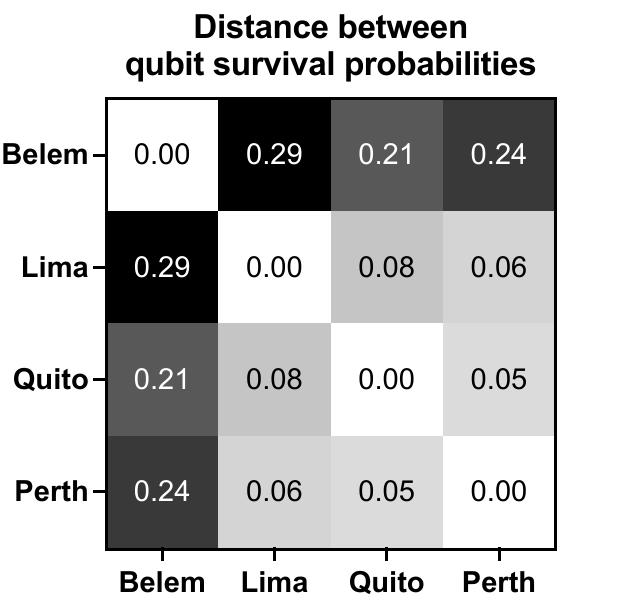}
    }
    \caption{Distinctiveness of noisy circuit execution results}
    \label{fig:circ_res_fp}
\end{figure}

First, we demonstrate that the noisy execution result of a circuit, influenced by the quantum operations involved, can serve as a distinctive identifier for quantum devices. An example is presented in Fig.~\ref{fig:prob_circ}. We construct a 3-qubit Bernstein-Vazirani (BV) circuit and transpile it with a qubit mapping of [0, 1, 3]. This circuit includes eight different quantum operations, some of which are repeated multiple times. It is executed on four different quantum devices, all of which share the same or a partially similar qubit topology, allowing the circuit to run on each. Subsequently, we calculate the Manhattan distance between each pair of execution results, as depicted in Fig.~\ref{fig:circ_res_fp} (a). This heatmap clearly illustrates that the execution results of the same circuit on different quantum devices vary significantly, underscoring their distinctiveness and quantifiability, thereby making them suitable as unique identifiers for each quantum machine.

\subsubsection{Dynamic fingerprint design.} 
\begin{figure}[h]
\centering 
\includegraphics[width=1\linewidth]{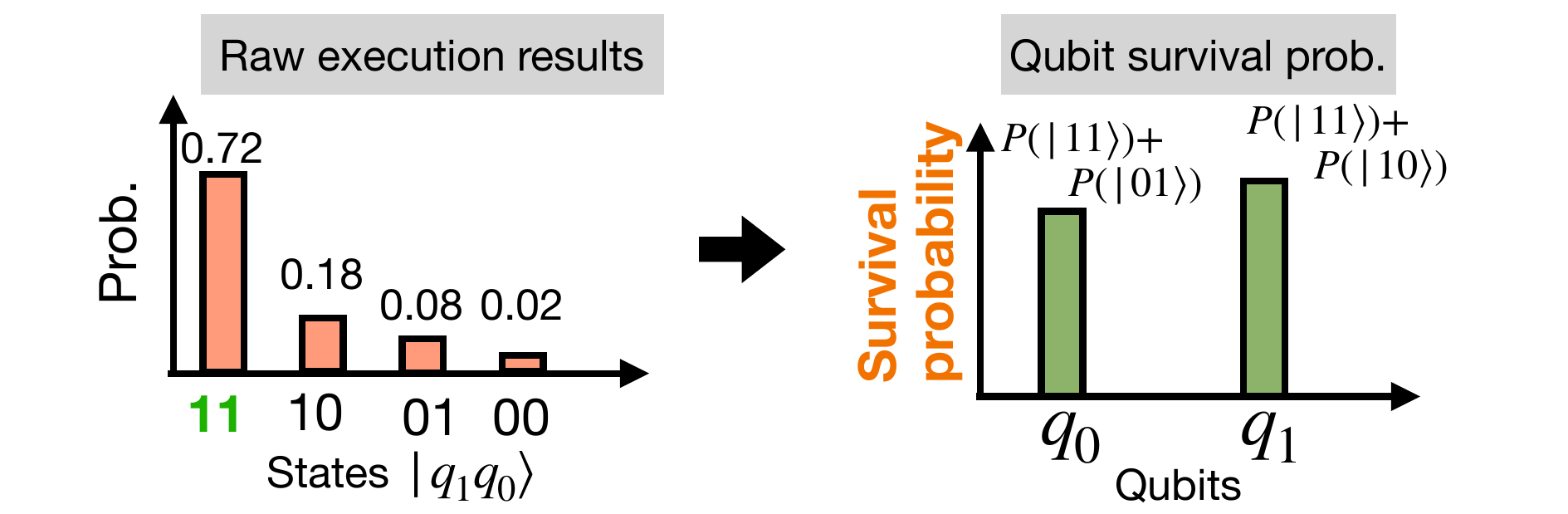}
\caption{Obtaining survival probability of qubits from the raw execution result.}
\label{fig:detive_qubit}
\end{figure}

To make it efficient for users to construct the user-side fingerprint, we do not use the raw execution results as the fingerprint of the device. Instead, we utilize the qubit survival probabilities of the probing circuit. The survival probability of a qubit refers to the probability of obtaining the correct state from that qubit.
The survival probability of a qubit is derived by calculating the marginal probability of the qubit remaining in the correct state from the generated distribution. Thus, the fingerprint is defined as \( F_d = [s_0, s_1, \ldots, s_n] \), where \( s_i \) represents the survival probability of the \( i \)-th measured qubit. 
For instance, as shown in Fig.~\ref{fig:detive_qubit}, the left side displays the raw execution results of a circuit.
Knowing that the correct state is $|11\rangle$, where each individual qubit's correct state is $|1\rangle$, the survival probability of qubit $q_0$ is calculated by $s_0 = P(|11\rangle) + P(|01\rangle)=0.8$. Similarly, the survival probability for qubit $q_1$ is $s_1 = P(|11\rangle) + P(|10\rangle)=0.9$. Thus, the device-side fingerprint is $F_d = [0.8, 0.9]$.
Note that this straightforward method of deriving qubit survival probabilities is only applicable to circuits that generate a single basis state, such as the BV circuit.
Fig.~\ref{fig:circ_res_fp} (b) demonstrates the effectiveness of using qubit survival probabilities as unique identifiers for quantum devices.

\subsubsection{Probing circuit} To ensure efficient and effective fingerprinting, the probing circuit must meet two criteria. First, it should generate a single basis state. Circuits with a single basis state result in a quantum system that lacks qubit state superpositions and inter-qubit entanglements after execution. This simplification facilitates the calculation of survival probabilities, which can be directly calculated from the raw execution results, as illustrated in Fig.~\ref{fig:detive_qubit}. Theoretically, in a quantum system without superposition and entanglement, the system's state is the tensor product of its individual qubits' states, allowing qubits to be considered independent. 
Secondly, the probing circuit should be both complex and small. It should involve enough noisy quantum operations to capture the error characteristics of the device, yet remain sufficiently small. This is because while the user only needs to estimate the survival probabilities of qubits considering operational errors, various errors can affect the noisy execution result. Therefore, a larger circuit could decrease the accuracy of user-side fingerprint generation, making it crucial that the circuit balances complexity with manageability. In our evaluation, we demonstrate that the 3-qubit BV circuit is an appropriate probing circuit.

\subsection{User-side Fingerprint} 

The user-side fingerprint represents the expected survival probability of the qubits in the probing circuit, assuming execution on the selected quantum device. To determine the user-side fingerprint, the user calculates these expected qubit survival probabilities by utilizing the operation errors recorded in the profile of the selected quantum device, as illustrated in step 2 of Fig.~\ref{fig:overview}. In this subsection, we introduce a qubit survival probability estimation algorithm that allows users to obtain the user-side fingerprint with linear complexity.

In essence, executing a quantum circuit entails the evolution of quantum information through a sequence of instructions, known as quantum state evolution. Unfortunately, quantum errors associated with noisy operations accumulate over time, resulting in a gradual decrease in the survival probability of the qubits. The output state of a circuit is determined by the states of its measured logical qubits, which pass through a series of noisy quantum gates and may require transfers between physical qubits. Consequently, our focus is on assessing the survival probability of these measured logical qubits. It has been shown that the amount of induced error is related to the involved gates, physical qubits, and qubit links \cite{nishio2020extracting}. To understand the effects of accumulated errors, we formulate a detailed gate-by-gate analysis that tracks the diminished survival probability of logical qubits across the quantum circuit. This survival probability of a qubit quantifies the likelihood that it maintains the correct quantum state after noisy operations. It is essential to note that we focus on the change in the survival probability and do not consider the exact state of the qubit during this analysis.

\textbf{Qubit survival probability estimation algorithm.} In this algorithm, each logical qubit can be considered individually due to the design of the probing circuit. For each qubit involved, we maintain a triple represented as \((q_i, Q_i, s_i)\), where \(q_i\) denotes the $i$-th logical qubit, \(Q_i\) is the location of \(q_i\), and \(s_i\) is the survival probability of \(q_i\). The location \(Q_i\) is initially set based on the logical-to-physical qubit mapping of the transpiled probing circuit and may be updated if \(q_i\) is swapped to another quantum register during circuit execution. The survival probability \(s_i\) starts at 1, assuming the state initialization is noiseless, which suggests that the quantum information in the qubits remains intact with 100\% probability prior to the execution of the probing circuit. As \(q_i\) passes through noisy quantum gates, its survival probability \(s_i\) decreases, with the rate of decrease dependent on the error rate of each encountered gate.

\begin{figure*}[t!]
\centering 
\includegraphics[width=1\textwidth]{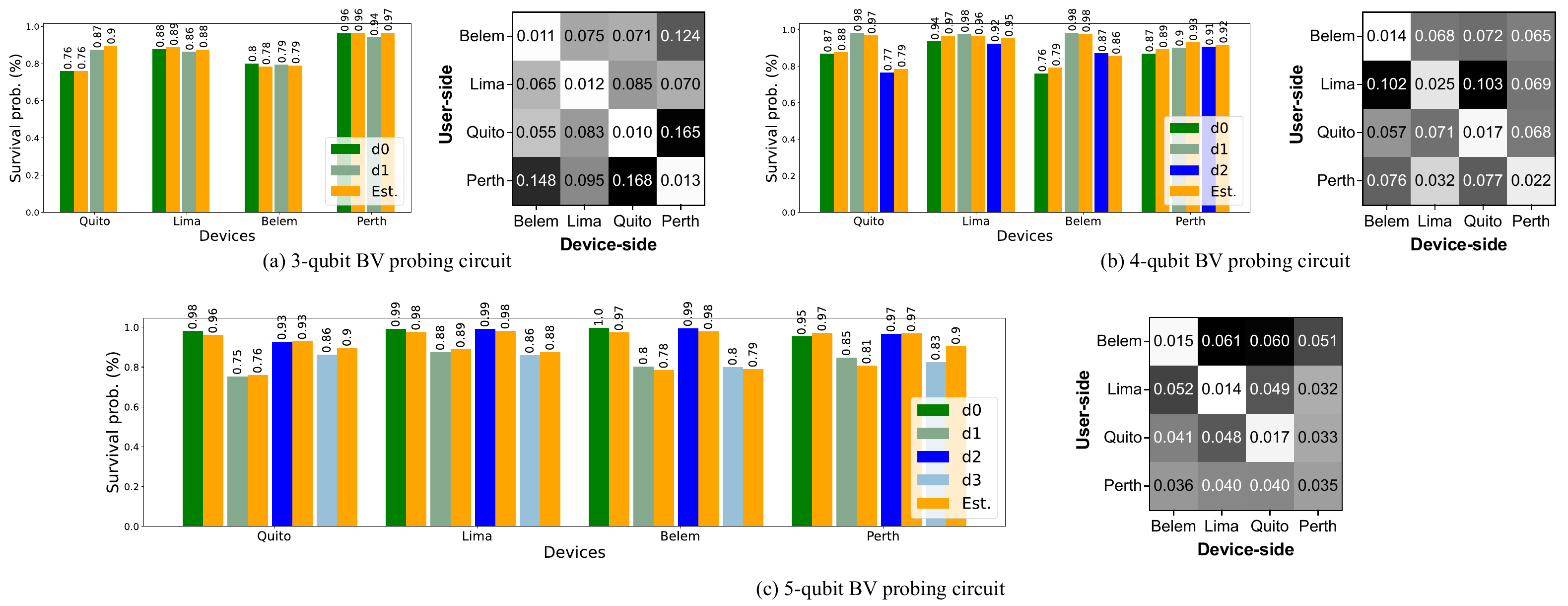}
\caption{Fingerprinting small size quantum devices.}
\label{fig:small_fp}
\end{figure*}

Consider a probing circuit represented as a sequence of quantum operations, denoted as $U = [Op_1,\ Op_2, \dots,\ Op_k]$. The $i$-th quantum operation $Op_i = (g_i,\ qreg_i,\ e_i)$ consists of three components: the noisy quantum gate $g_i$, the list of involved physical quantum registers $qreg_i$, and the error rate $e_i$ associated with the gate that is retrieved from the error profile of the selected quantum device.
When executing the quantum operation $Op_i = (g_i,\ qreg_i,\ e_i)$, two possible cases arise for performing the survival probability reduction on the involved logical qubits. In the first case, if $g_i$ is a state modification gate that modifies the quantum information stored in the physical qubits $qreg_i$, or qubit measurement, the survival probability of $q_i$ such that $Q_i \in qreg_i$ will be updated as follows:
\begin{equation}
s_i = s_i \times (1 - e_i)
\end{equation}
In the second case, when the gate $g_i$ is a SWAP gate that switches the location of two qubits $q_x$ and $q_y$, their survival probabilities will be updated as $s_x = s_x \times (1 - e_i) ^ 3$ and $s_y = s_y \times (1 - e_i) ^ 3$, respectively.
Since the SWAP gate consists of three CNOT gates, the survival probabilities of $s_x$ and $s_y$ are modified accordingly. Additionally, the locations of the logical qubits will be updated by exchanging the values of $Q_x$ and $Q_y$.

For example, as depicted in Fig.~\ref{fig:overview} step 2, the qubit $q_1$ is initially assigned to the physical qubit $Q_1$ and encounters an H gate with an error rate of 0.0015. It is subsequently transferred to $Q_3$ via a SWAP gate (comprised of three CNOT gates, each with an error rate of 0.013), leading to a reduction in its survival probability to $s_1 = 1 \times (1-0.0015) \times (1-0.013)^3 = 0.96$. It then traverses a CNOT gate with the same error rate, resulting in $s_1 = 0.96 \times (1-0.013) = 0.9472$. Following this, it passes through another H gate with an error rate of 0.0004, further reducing its survival probability to $s_1 = 0.9472 \times (1-0.0004) = 0.9468$. Ultimately, it is measured on $Q_3$ with an error rate of 0.042, bringing its final survival probability $s_i$ down to approximately $0.9468 \times (1-0.042) \approx 0.91$.

\section{Evaluation} \label{sec:eval}

We assess the proposed dynamic fingerprinting approach for detecting fraudulent service provision on the quantum cloud platform.
Our experiments were conducted on the IBM quantum cloud platform using Qiskit \cite{wille2019ibm}, though the approach is applicable to any gate-based quantum cloud platform.
The quantum devices used in our experiments were Quito ($ibmq\_quito$), Lima ($ibmq\_lima$), Belem ($ibmq\_belem$), and Perth ($ibm\_perth$). Quito, Lima, and Belem each have five qubits with identical qubit topologies, while Perth contains seven qubits with a topology that partially overlaps with the others. We also included experiments on the recently released 127-qubit machines: Brisbane ($ibm\_brisbane$), Osaka ($ibm\_osaka$), and Kyoto ($ibm\_kyoto$).
To alleviate the impact of fluctuations in the quality of quantum hardware, each quantum circuit was executed in three rounds, with each round consisting of 4000 shots.

\begin{figure*}[t!]
\centering 
\includegraphics[width=0.95\textwidth]{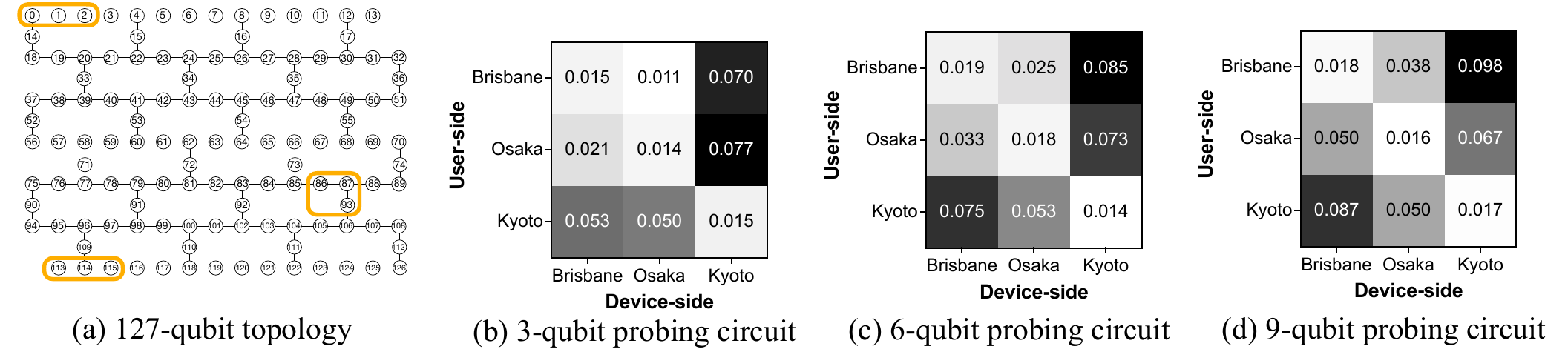}
\caption{Fingerprinting large size quantum devices.
}
\label{fig:large_fp}
\end{figure*}

\subsection{Accuracy of Fingerprinting}
We assess the performance of dynamic fingerprinting in identifying quantum devices.
We focus on identifying devices among similarly sized quantum devices, as their identical or partially identical qubit topologies allow for substitution between them. In contrast, quantum devices with significantly different sizes, such as a 5-qubit machine compared to a 127-qubit machine, exhibit distinct qubit topologies. Consequently, a circuit transpiled for one machine may not be executable on another due to these topological differences, resulting in the cloud platform reporting an error. In such cases, devices reporting errors are immediately considered unmatched.
Therefore, we categorize quantum devices into two groups. The first group comprises small-sized devices: Quito, Lima, Belem, and Perth. The second group consists of relatively large-sized devices: Brisbane, Osaka, and Kyoto. The results for the two groups are depicted in Fig.~\ref{fig:small_fp} for the first group and Fig.~\ref{fig:large_fp} for the second group, respectively.

\subsubsection{Small-size devices}
We use three different sizes of BV circuits as probing circuits. For each BV circuit constructed with \( n \) qubits, where \( n-1 \) qubits are measured, the fingerprint is a vector of size \( n-1 \).
In the subfigures of Fig.~\ref{fig:small_fp}, the left panel illustrates the device-side and user-side fingerprints. The bar labeled \( d_i \) represents the survival probability of the \( i \)-th qubit based on cloud execution results, while the adjacent orange bar, labeled \( Est. \), shows the user-side calculated survival probability of \( d_i \) using the proposed algorithm.

Firstly, the subfigures show significant variation in qubit survival probabilities for the same circuit across different quantum devices, highlighting the effectiveness of using qubit survival probability as a distinctive identifier. Secondly, the close alignment between the device-side and user-side fingerprints of the same device illustrates the accuracy of our method in capturing qubit survival probabilities. On average, the qubit-wise differences between the device-side and user-side fingerprints for the same devices are 0.0115, 0.0195, and 0.0203 for the three probing circuit sizes, respectively. The increased disparity suggests that the accuracy of the proposed survival probability estimation algorithm may decline for larger probing circuits due to the impact of crosstalk and decoherence errors. Therefore, a probing circuit with 3 or 4 qubits is preferred.

In the right panels of Fig.~\ref{fig:small_fp}, the qubit-wise Manhattan distances between the device-side and user-side fingerprints of various quantum device pairs are displayed. 
In the tables, each row represents a device identification result.
For instance, in the right panel of Fig. \ref{fig:small_fp} (a), the first row illustrates how the user identifies the Belem machine from four unknown quantum devices. The user-side fingerprint of Belem is compared with the device-side fingerprints of all four devices. The first cell, which shows the smallest Manhattan distance, indicates that the user identifies the first device as Belem, accurately matching the device's actual identity. In summary, the results demonstrate that the user can accurately identify the quantum device in 11 out of 12 fingerprinting experiments.

\begin{figure}[t]
\centering 
\includegraphics[width=0.9\linewidth]{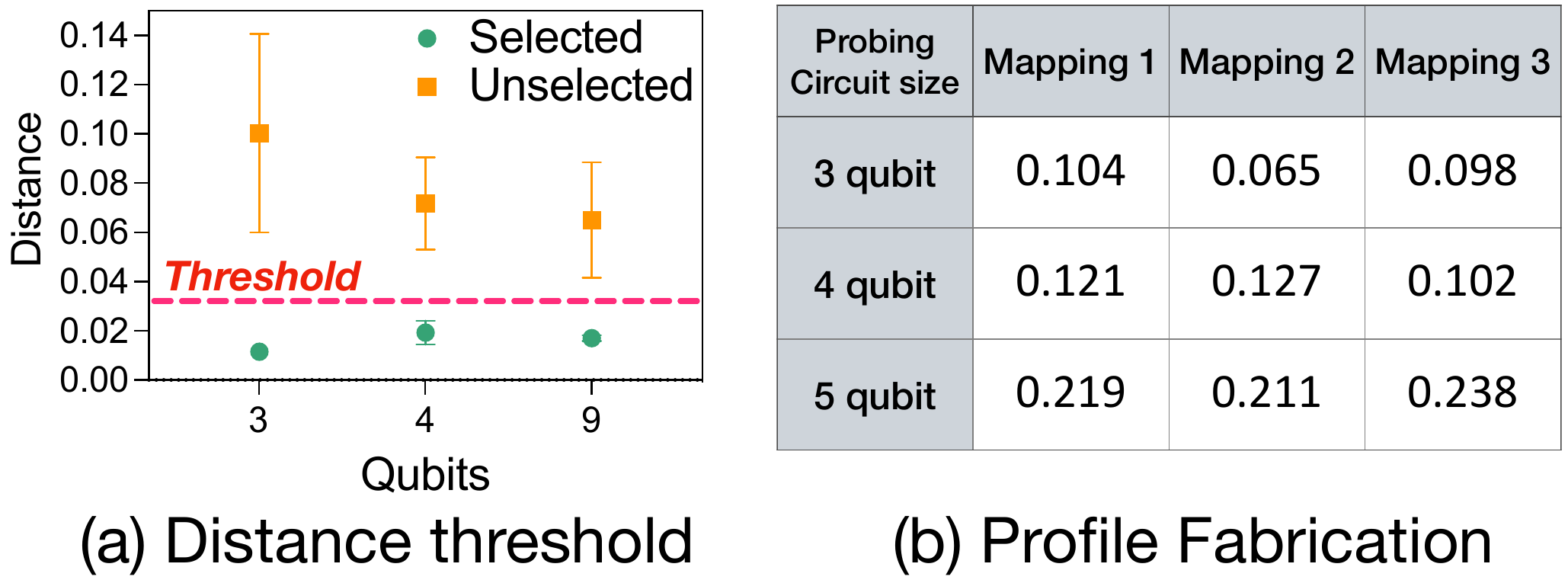}
\caption{ (a) Determining the fingerprint distance threshold. (b) An example of a profile fabrication attack. }
\label{fig:tolerance}
\end{figure}

\subsubsection{Large-size devices}
 The 127-qubit quantum devices Brisbane, Osaka, and Kyoto share the same qubit topology as depicted in Fig.~\ref{fig:large_fp} (a). To ensure accurate user-side fingerprint generation, rather than constructing an integrated, large probing circuit, we choose to build a probing circuit comprising several independent subcircuits.
The subcircuits are spaced sufficiently apart to minimize the risk of crosstalk errors. In this experiment, we deployed three such 3-qubit probing subcircuits on the quantum device, as shown in Fig.~\ref{fig:large_fp} (a). 

Fig.~\ref{fig:large_fp} (b) through (d) illustrate the fingerprinting performance of three quantum devices using probing circuits of varying sizes. 
Apart from the first experiment (first row) shown in Fig. \ref{fig:large_fp} (b), all remaining eight fingerprinting trials correctly identify the quantum device. 
Fingerprinting on 127-qubit devices demonstrates that a 6-qubit probing circuit, composed of two 3-qubit subcircuits, is sufficient to identify quantum devices. However, a 9-qubit probing circuit, with three 3-qubit subcircuits, creates a distinct gap that makes it easier to identify the best-matching device. Thus, we conclude that a 9-qubit probing circuit is sufficient to confidently fingerprint devices, underscoring the scalability of the proposed fingerprinting approach.

\subsection{Detection of Fraudulent Service}

We now evaluate the effectiveness of detecting fraudulent services. To achieve successful detection with a single probing circuit, users need to establish a threshold for differences between device-side and user-side fingerprints. Therefore, we analyze the fingerprinting results to establish the threshold by considering probing circuits with 3, 4, and 9 qubits, which are executed on devices containing 5, 7, or 127 qubits. Each circuit is transpiled using three different qubit mappings. Fig. \ref{fig:tolerance} (a) illustrates the distance between device-side and user-side fingerprints for selected and unselected devices. Notably, for all three probing circuit sizes, there is a distinct gap between these two sets of differences. This observation allows us to establish a difference threshold of 0.035. If the Manhattan distance between fingerprints surpasses 0.035, it indicates the presence of a fraudulent service.

Furthermore, we detect an actual profile fabrication attack on the Belem device. In Fig. \ref{fig:tolerance} (b), the Manhattan distances between the device-side and user-side fingerprints for Belem are displayed, which includes probing circuits of three different sizes, and each is transpiled using three distinct qubit mappings.
Ideally, all distances should be less than the previously determined threshold of 0.035. However, all distances significantly exceed this threshold, indicating that the quantum device deviates significantly from the performance recorded in its profile. This discrepancy is due to the device nearing retirement, resulting in their insufficient calibration and outdated profile information. 
It is not a case of the IBM Quantum cloud provider intentionally fabricating the profile to attack users, but it demonstrates the effectiveness of detecting profile fabrication attacks.

\section{Conclusion} \label{sec:concl}
We propose an approach for detecting fraudulent services on quantum cloud platforms through dynamic fingerprinting, which leverages the diversity of quantum errors without being hindered by their instability. Our results demonstrate that the noisy execution results (qubit survival probabilities) of a single probing circuit can uniquely identify quantum devices. We introduce an efficient algorithm for users to construct user-side fingerprints.
With the proposed dynamic fingerprinting approach, only a single probing circuit is required for each detection. This trivial cost makes the method suitable for current quantum cloud platforms, where user access is both costly and limited.
Our experiments on the IBM Quantum cloud platform confirm the accuracy of this fingerprinting approach, revealing that setting the fingerprint distance threshold at 0.035 allows the method to accurately detect fraudulent services.

\begin{acks}
The authors would like to thank all the reviewers for their helpful comments. 
This work was supported by the Commonwealth Cyber Initiative (cyberinitiative.org).
\end{acks}

% \bibliographystyle{ACM-Reference-Format}
% \bibliography{references.bib}

\bibliographystyle{ACM-Reference-Format}
\bibliography{0_ref}


\begin{thebibliography}{24}

%%% the Web address.

\ifx \showCODEN    \undefined \def \showCODEN     #1{\unskip}     \fi
\ifx \showDOI      \undefined \def \showDOI       #1{#1}\fi
\ifx \showISBNx    \undefined \def \showISBNx     #1{\unskip}     \fi
\ifx \showISBNxiii \undefined \def \showISBNxiii  #1{\unskip}     \fi
\ifx \showISSN     \undefined \def \showISSN      #1{\unskip}     \fi
\ifx \showLCCN     \undefined \def \showLCCN      #1{\unskip}     \fi
\ifx \shownote     \undefined \def \shownote      #1{#1}          \fi
\ifx \showarticletitle \undefined \def \showarticletitle #1{#1}   \fi
\ifx \showURL      \undefined \def \showURL       {\relax}        \fi
% invisible to TeX
\providecommand\bibfield[2]{#2}
\providecommand\bibinfo[2]{#2}
\providecommand\natexlab[1]{#1}
\providecommand\showeprint[2][]{arXiv:#2}

\bibitem[Ama(2023)]%
        {AmazonBraket}
 \bibinfo{year}{2023}\natexlab{}.
\newblock \bibinfo{title}{Amazon Braket}.
\newblock
\newblock
\urldef\tempurl%
\url{https://aws.amazon.com/braket/}
\showURL{%
\tempurl}


\bibitem[IBM(2023)]%
        {IBMQ}
 \bibinfo{year}{2023}\natexlab{}.
\newblock \bibinfo{title}{Now entering the era of quantum utility}.
\newblock
\newblock
\urldef\tempurl%
\url{https://www.ibm.com/quantum}
\showURL{%
\tempurl}


\bibitem[Azu(2023)]%
        {Azure}
 \bibinfo{year}{2023}\natexlab{}.
\newblock \bibinfo{title}{What is Azure Quantum?}
\newblock
\newblock
\urldef\tempurl%
\url{https://learn.microsoft.com/en-us/azure/quantum/overview-azure-quantum}
\showURL{%
\tempurl}


\bibitem[Ash-Saki et~al\mbox{.}(2020)]%
        {ash2020analysis}
\bibfield{author}{\bibinfo{person}{Abdullah Ash-Saki}, \bibinfo{person}{Mahabubul Alam}, {and} \bibinfo{person}{Swaroop Ghosh}.} \bibinfo{year}{2020}\natexlab{}.
\newblock \showarticletitle{Analysis of crosstalk in NISQ devices and security implications in multi-programming regime}. In \bibinfo{booktitle}{\emph{Proceedings of the ACM/IEEE International Symposium on Low Power Electronics and Design}}. \bibinfo{pages}{25--30}.
\newblock


\bibitem[Castelvecchi(2017)]%
        {castelvecchi2017ibm}
\bibfield{author}{\bibinfo{person}{Davide Castelvecchi}.} \bibinfo{year}{2017}\natexlab{}.
\newblock \showarticletitle{IBM's quantum cloud computer goes commercial}.
\newblock \bibinfo{journal}{\emph{Nature}} \bibinfo{volume}{543}, \bibinfo{number}{7644} (\bibinfo{year}{2017}).
\newblock


\bibitem[Das et~al\mbox{.}(2019)]%
        {das2019case}
\bibfield{author}{\bibinfo{person}{Poulami Das}, \bibinfo{person}{Swamit~S Tannu}, \bibinfo{person}{Prashant~J Nair}, {and} \bibinfo{person}{Moinuddin Qureshi}.} \bibinfo{year}{2019}\natexlab{}.
\newblock \showarticletitle{A case for multi-programming quantum computers}. In \bibinfo{booktitle}{\emph{Proceedings of the 52nd Annual IEEE/ACM International Symposium on Microarchitecture}}. \bibinfo{pages}{291--303}.
\newblock


\bibitem[Knill et~al\mbox{.}(2008)]%
        {knill2008randomized}
\bibfield{author}{\bibinfo{person}{Emanuel Knill}, \bibinfo{person}{Dietrich Leibfried}, \bibinfo{person}{Rolf Reichle}, \bibinfo{person}{Joe Britton}, \bibinfo{person}{R~Brad Blakestad}, \bibinfo{person}{John~D Jost}, \bibinfo{person}{Chris Langer}, \bibinfo{person}{Roee Ozeri}, \bibinfo{person}{Signe Seidelin}, {and} \bibinfo{person}{David~J Wineland}.} \bibinfo{year}{2008}\natexlab{}.
\newblock \showarticletitle{Randomized benchmarking of quantum gates}.
\newblock \bibinfo{journal}{\emph{Physical Review A}} \bibinfo{volume}{77}, \bibinfo{number}{1} (\bibinfo{year}{2008}), \bibinfo{pages}{012307}.
\newblock


\bibitem[Magesan et~al\mbox{.}({[n.\,d.]})]%
        {magesan3639robust}
\bibfield{author}{\bibinfo{person}{E Magesan}, \bibinfo{person}{JM Gambetta}, {and} \bibinfo{person}{J Emerson}.} \bibinfo{year}{[n.\,d.]}\natexlab{}.
\newblock \showarticletitle{Robust randomized benchmarking of quantum processes}.
\newblock \bibinfo{journal}{\emph{arXiv preprint arXiv:1009.3639}} (\bibinfo{year}{[n.\,d.]}).
\newblock


\bibitem[Magesan et~al\mbox{.}(2012)]%
        {magesan2012characterizing}
\bibfield{author}{\bibinfo{person}{Easwar Magesan}, \bibinfo{person}{Jay~M Gambetta}, {and} \bibinfo{person}{Joseph Emerson}.} \bibinfo{year}{2012}\natexlab{}.
\newblock \showarticletitle{Characterizing quantum gates via randomized benchmarking}.
\newblock \bibinfo{journal}{\emph{Physical Review A}} \bibinfo{volume}{85}, \bibinfo{number}{4} (\bibinfo{year}{2012}), \bibinfo{pages}{042311}.
\newblock


\bibitem[Martina et~al\mbox{.}(2022)]%
        {martina2022learning}
\bibfield{author}{\bibinfo{person}{Stefano Martina}, \bibinfo{person}{Lorenzo Buffoni}, \bibinfo{person}{Stefano Gherardini}, {and} \bibinfo{person}{Filippo Caruso}.} \bibinfo{year}{2022}\natexlab{}.
\newblock \showarticletitle{Learning the noise fingerprint of quantum devices}.
\newblock \bibinfo{journal}{\emph{Quantum Machine Intelligence}} \bibinfo{volume}{4}, \bibinfo{number}{1} (\bibinfo{year}{2022}), \bibinfo{pages}{8}.
\newblock


\bibitem[Mi et~al\mbox{.}(2021)]%
        {mi2021short}
\bibfield{author}{\bibinfo{person}{Allen Mi}, \bibinfo{person}{Shuwen Deng}, {and} \bibinfo{person}{Jakub Szefer}.} \bibinfo{year}{2021}\natexlab{}.
\newblock \showarticletitle{Short paper: Device-and locality-specific fingerprinting of shared nisq quantum computers}. In \bibinfo{booktitle}{\emph{Workshop on Hardware and Architectural Support for Security and Privacy}}. \bibinfo{pages}{1--6}.
\newblock


\bibitem[Mi et~al\mbox{.}(2022)]%
        {mi2022securing}
\bibfield{author}{\bibinfo{person}{Allen Mi}, \bibinfo{person}{Shuwen Deng}, {and} \bibinfo{person}{Jakub Szefer}.} \bibinfo{year}{2022}\natexlab{}.
\newblock \showarticletitle{Securing Reset Operations in NISQ Quantum Computers}. In \bibinfo{booktitle}{\emph{Proceedings of the 2022 ACM SIGSAC Conference on Computer and Communications Security}}. \bibinfo{pages}{2279--2293}.
\newblock


\bibitem[Morris et~al\mbox{.}(2023)]%
        {morris2023fingerprinting}
\bibfield{author}{\bibinfo{person}{Jalil Morris}, \bibinfo{person}{Anisul Abedin}, \bibinfo{person}{Chuanqi Xu}, {and} \bibinfo{person}{Jakub Szefer}.} \bibinfo{year}{2023}\natexlab{}.
\newblock \showarticletitle{Fingerprinting Quantum Computer Equipment}. In \bibinfo{booktitle}{\emph{Proceedings of the Great Lakes Symposium on VLSI 2023}}. \bibinfo{pages}{117--123}.
\newblock


\bibitem[Nishio et~al\mbox{.}(2020)]%
        {nishio2020extracting}
\bibfield{author}{\bibinfo{person}{Shin Nishio}, \bibinfo{person}{Yulu Pan}, \bibinfo{person}{Takahiko Satoh}, \bibinfo{person}{Hideharu Amano}, {and} \bibinfo{person}{Rodney~Van Meter}.} \bibinfo{year}{2020}\natexlab{}.
\newblock \showarticletitle{Extracting success from ibm’s 20-qubit machines using error-aware compilation}.
\newblock \bibinfo{journal}{\emph{ACM Journal on Emerging Technologies in Computing Systems (JETC)}} \bibinfo{volume}{16}, \bibinfo{number}{3} (\bibinfo{year}{2020}), \bibinfo{pages}{1--25}.
\newblock


\bibitem[Phalak et~al\mbox{.}(2021)]%
        {phalak2021quantum}
\bibfield{author}{\bibinfo{person}{Koustubh Phalak}, \bibinfo{person}{Abdullah Ash-Saki}, \bibinfo{person}{Mahabubul Alam}, \bibinfo{person}{Rasit~Onur Topaloglu}, {and} \bibinfo{person}{Swaroop Ghosh}.} \bibinfo{year}{2021}\natexlab{}.
\newblock \showarticletitle{Quantum puf for security and trust in quantum computing}.
\newblock \bibinfo{journal}{\emph{IEEE Journal on Emerging and Selected Topics in Circuits and Systems}} \bibinfo{volume}{11}, \bibinfo{number}{2} (\bibinfo{year}{2021}), \bibinfo{pages}{333--342}.
\newblock


\bibitem[Preskill(2018)]%
        {preskill2018quantum}
\bibfield{author}{\bibinfo{person}{John Preskill}.} \bibinfo{year}{2018}\natexlab{}.
\newblock \showarticletitle{Quantum computing in the NISQ era and beyond}.
\newblock \bibinfo{journal}{\emph{Quantum}}  \bibinfo{volume}{2} (\bibinfo{year}{2018}), \bibinfo{pages}{79}.
\newblock


\bibitem[Rahaman and Islam(2015)]%
        {rahaman2015review}
\bibfield{author}{\bibinfo{person}{Mijanur Rahaman} {and} \bibinfo{person}{Md~Masudul Islam}.} \bibinfo{year}{2015}\natexlab{}.
\newblock \showarticletitle{A review on progress and problems of quantum computing as a service (QcaaS) in the perspective of cloud computing}.
\newblock \bibinfo{journal}{\emph{Global Journal of Computer Science and Technology}} \bibinfo{volume}{15}, \bibinfo{number}{4} (\bibinfo{year}{2015}).
\newblock


\bibitem[Ravi et~al\mbox{.}(2021)]%
        {ravi2021quantum}
\bibfield{author}{\bibinfo{person}{Gokul~Subramanian Ravi}, \bibinfo{person}{Kaitlin~N Smith}, \bibinfo{person}{Pranav Gokhale}, {and} \bibinfo{person}{Frederic~T Chong}.} \bibinfo{year}{2021}\natexlab{}.
\newblock \showarticletitle{Quantum computing in the cloud: Analyzing job and machine characteristics}. In \bibinfo{booktitle}{\emph{2021 IEEE International Symposium on Workload Characterization (IISWC)}}. IEEE, \bibinfo{pages}{39--50}.
\newblock


\bibitem[Singh and Sachdev(2014)]%
        {singh2014quantum}
\bibfield{author}{\bibinfo{person}{Harpreet Singh} {and} \bibinfo{person}{Abha Sachdev}.} \bibinfo{year}{2014}\natexlab{}.
\newblock \showarticletitle{The quantum way of cloud computing}. In \bibinfo{booktitle}{\emph{2014 International Conference on Reliability Optimization and Information Technology (ICROIT)}}. Ieee, \bibinfo{pages}{397--400}.
\newblock


\bibitem[Smith et~al\mbox{.}(2023)]%
        {smith2023fast}
\bibfield{author}{\bibinfo{person}{Kaitlin~N Smith}, \bibinfo{person}{Joshua Viszlai}, \bibinfo{person}{Lennart~Maximilian Seifert}, \bibinfo{person}{Jonathan~M Baker}, \bibinfo{person}{Jakub Szefer}, {and} \bibinfo{person}{Frederic~T Chong}.} \bibinfo{year}{2023}\natexlab{}.
\newblock \showarticletitle{Fast Fingerprinting of Cloud-based NISQ Quantum Computers}. In \bibinfo{booktitle}{\emph{2023 IEEE International Symposium on Hardware Oriented Security and Trust (HOST)}}. IEEE, \bibinfo{pages}{1--12}.
\newblock


\bibitem[Upadhyay and Ghosh(2023)]%
        {upadhyay2023obfuscating}
\bibfield{author}{\bibinfo{person}{Suryansh Upadhyay} {and} \bibinfo{person}{Swaroop Ghosh}.} \bibinfo{year}{2023}\natexlab{}.
\newblock \showarticletitle{Obfuscating Quantum Hybrid-Classical Algorithms for Security and Privacy}.
\newblock \bibinfo{journal}{\emph{arXiv preprint arXiv:2305.02379}} (\bibinfo{year}{2023}).
\newblock


\bibitem[Upadhyay et~al\mbox{.}(2023)]%
        {upadhyay2023trustworthy}
\bibfield{author}{\bibinfo{person}{Suryansh Upadhyay}, \bibinfo{person}{Rasit~Onur Topaloglu}, {and} \bibinfo{person}{Swaroop Ghosh}.} \bibinfo{year}{2023}\natexlab{}.
\newblock \showarticletitle{Trustworthy Computing using Untrusted Cloud-Based Quantum Hardware}.
\newblock \bibinfo{journal}{\emph{arXiv preprint arXiv:2305.01826}} (\bibinfo{year}{2023}).
\newblock


\bibitem[Wille et~al\mbox{.}(2019)]%
        {wille2019ibm}
\bibfield{author}{\bibinfo{person}{Robert Wille}, \bibinfo{person}{Rod Van~Meter}, {and} \bibinfo{person}{Yehuda Naveh}.} \bibinfo{year}{2019}\natexlab{}.
\newblock \showarticletitle{IBM’s Qiskit tool chain: Working with and developing for real quantum computers}. In \bibinfo{booktitle}{\emph{2019 Design, Automation \& Test in Europe Conference \& Exhibition (DATE)}}. IEEE, \bibinfo{pages}{1234--1240}.
\newblock


\bibitem[Xu et~al\mbox{.}(2023)]%
        {xu2023exploration}
\bibfield{author}{\bibinfo{person}{Chuanqi Xu}, \bibinfo{person}{Ferhat Erata}, {and} \bibinfo{person}{Jakub Szefer}.} \bibinfo{year}{2023}\natexlab{}.
\newblock \showarticletitle{Exploration of Quantum Computer Power Side-Channels}.
\newblock \bibinfo{journal}{\emph{arXiv preprint arXiv:2304.03315}} (\bibinfo{year}{2023}).
\newblock


\end{thebibliography}

\end{document}